\documentclass[
    ,final            
  ]
  {aipproc}

\layoutstyle{8x11double}

\newcommand{\ba}{\begin{array}}
\newcommand{\ea}{\end{array}}
\newcommand{\bd}{\begin{displaymath}}
\newcommand{\ed}{\end{displaymath}}
\newcommand{\be}{\begin{equation}}
\newcommand{\ee}{\end{equation}}
\newcommand{\bea}{\begin{eqnarray}}
\newcommand{\eea}{\end{eqnarray}}

\def\a{\alpha}

\def\b{\beta}

\def\q2 {q^2}

\def\bt{\begin{table}}
\def\et{\end{table}}

\begin{document}

\title{Signatures of non-universal gaugino and scalar masses 
at the Large Hadron Collider}

\classification{12.60.Jv}
\keywords      {Supersymmetry, Supersymmetric Models
\\{\bf{Preprint}}: HRI-P-08-09-001,HRI-RECAPP-2008-011} 

\author{Subhaditya Bhattacharya}{
  address={Regional Centre for Accelerator-based Particle Physics \\
     Harish-Chandra Research Institute\\
Chhatnag Road, Jhunsi, Allahabad - 211 019, India\\
e-mail: subha@hri.res.in} 
}
\begin{abstract}
We perform a multichannel analysis in 
context of the Large Hadron Collider (LHC) 
for supersymmetric (SUSY) theories with high-scale 
non-universal gaugino masses arising from different non-singlet 
representations of $SU(5)$ and $SO(10)$ gauge groups in a SUSY-GUT scenario 
and non-universal scalar masses in form of squark-slepton 
non-universality, third family scalar non-universality and that 
arising due to $SO(10)$ $D$-terms. 
We present the numerical predictions over a wide region of parameter space 
using event generator {\tt Pythia}. Certain broad features emerge from 
the study which may be useful to identify these non-universal schemes 
and distinguish them from the minimal supergravity (mSUGRA) framework.
\end{abstract}

\maketitle

\paragraph{{\bf{Introduction}:}}
As the LHC takes off at CERN, the search for SUSY at the TeV 
scale\cite{Book} enters into an exciting era. Some of the popular SUSY
scenarios are motivated by the so-called minimal supergravity (mSUGRA)
\cite{mSUGRA} framework with universal gaugino 
and scalar masses at the unification scale. However, 
{\it there is scope for non-universality in the 
gaugino and scalar masses}, a feature that can 
often have crucial impact on the 
experimental signatures. In a recent work \cite{Subho1}, we performed a 
multichannel analysis for SUSY theories with non-universal gaugino mass, 
arising from the non-singlet representations 
of $SU(5)$ and $SO(10)$ GUT groups 
followed by another one \cite{subho2} where we analyzed similarly 
scalar non-universality arising from squark-slepton non-universality, 
third family scalar non-universality and non-universality coming from 
$SO(10)~D$-terms in context of the LHC. Such systematic 
analyses, based on various observable signals at the LHC, could indicate a 
departure from universality. In both the studies we present our 
results in terms of the ratios of two different signals, which to our 
advantage reduces uncertainties due to parton distributions, 
factorization scales, jet energy resolutions etc.

\paragraph{{\bf{Non-universal gaugino masses}:}}
Non-universality in gaugino masses arise as the non-trivial 
gauge kinetic function $f_{\a  \b}(\Phi^{j})$ can be expanded in terms 
of the non-singlet components of the hidden sector scalar fields $\Phi^N$ 
\cite{Ellis}:
\vspace {-0.1cm}
\bea
f_{\a \b}(\Phi^{j})= f_{0}(\Phi^{S})\delta_{\a
  \b}+\sum_{N}\xi_{N}(\Phi^s)
\frac{{\Phi^{N}}_{\a \b}}{M}+ {\mathcal{O}}(\frac{\Phi^N}{M})^2
\eea

\vspace {-0.1cm}
Non-singlet representations to which $\Phi^{N}$ can belong:

\noindent 
For $SU(5)$:

$(24\times 24)_{symm}=1+24+75+200$

\noindent
For $SO(10)$:

$(45\times 45)_{symm}=1+54+210+770$ 

\noindent
\vspace {0.05cm}

Once these $\Phi^{N}$s get VEVs, $f_{\a  \b}$ and gaugino masses
become non-universal.
\vspace {-0.2cm}
\begin{center}

\begin{tabular}{|c|c|}

\multicolumn{2}{c}{Table 1: Gaugino mass ratios for SU(5). }\\
\hline
 Representation & $M_{3}:M_{2}:M_{1}$ at $M_{GUT}$ \\
\hline 
{\bf 1} & 1:1:1 \\
\hline
{\bf 24} & 2:(-3):(-1) \\
\hline
{\bf 75} & 1:3:(-5) \\
\hline
{\bf 200} & 1:2:10 \\
\hline
\end {tabular}\\

\noindent
\end {center}

\noindent

\vspace {-0.2cm}
\begin{center}

\begin{tabular}{|c|c|}

\multicolumn{2}{c}{Table 2: Gaugino mass ratios for SO(10).}\\ 
\multicolumn{2}{c}{(Only for the lowest non-singlet representation)}\\
\hline
 Representation & $M_{3}:M_{2}:M_{1}$ at $M_{G}$ \\
\hline
{ 1} & 1:1:1 \\
\hline
{\bf 54(i)}: {$H \rightarrow SU(2) \times SO(7)$} & 1:(-7/3):1 \\
\hline
{\bf 54(ii)}: {$H \rightarrow SU(4) \times SU(2) \times SU(2)$} & 
1:(-3/2):(-1) \\
\hline
\end {tabular}\\
\noindent

\end {center}

Here one assumes that the breaking of the GUT group to the Standard Model (SM) 
takes place at the GUT scale itself, 
without any intermediate scale being there.
\paragraph{{\bf{Non-universal scalar masses}:}}
Here we study three models as follows:\\
(i) {\bf{Squark-slepton Non-universality}:} Here the squarks and slepton 
masses at low-energy are results of evolution from
mutually uncorrelated mass parameters 
($m_{0\tilde{q}}$ and $m_{0{\tilde l}}$ respectively) at a high scale.
Although this is a purely phenomenological approach,
it is helpful in the sense that it embodies the complete
independence of the coloured and uncloured 
scalar masses at the high scale. 
(ii) {\bf{Non-universality in the third family}:} Scalar masses in the third 
family evolve from a separate high-scale mass parameter $m^3_0$, 
while a different parameter $m^{(1,2)}_0$ is the origin of scalar
masses in the first two families. This is motivated from string-inspired 
models, which also addresses FCNC problems with `inverted hierarchy' 
\cite{3rd1}.
(iii) {\bf{Non-universality due to $SO(10) ~D$-terms}:}In an $SO(10)$ 
framework, the matter fields belong to the representation {\bf 16}, and 
can be further classified into sub-multiplets, depending 
on the representations of $SU(5)$ to which they belong. 
In this classification, expressing the (s)fermions generically 
to include all families, the superfields $D^c$ and $L$  belong 
to $\bar{\bf 5}$, while $Q$, $U^c$ and $E^c$ belong to {\bf 10}.
These acquire different $D$-terms when $SO(10)$ breaks to 
SM gauge group\cite{so102}:
\vspace{-0.1 cm}
\bea
m^2_{\bar 5} = m^2_0 - 1.5 D m^2_0 ~~~ (for~ D^c~ \&~ L) \\
m^2_{10} = m^2_0 + 0.5 D m^2_0  ~~~(for~ E^c, U^c~  \&~ Q)
\eea
\noindent
Here $D$ is a dimensionless parameter quantifying the
added contribution to the SUSY breaking terms in terms of 
the `universal' high-scale mass parameter $m_0$.

\paragraph{{\bf{Strategy for simulation}:}}
The spectrum generated by {\tt SuSpect} v2.3 \cite{SuSpect}
in each of theses scenarios is fed into the event generator {\tt Pythia} 
6.405 \cite{PYTHIA} for the simulation of $pp$ 
collision with centre of mass energy 14 TeV.
We have used {\tt CTEQ5L} \cite{CTEQ} parton distribution functions, 
the QCD renormalization and factorization scales being
both set at the subprocess centre-of-mass energy  $\sqrt{\hat{s}}$. All 
possible SUSY processes with conserved $R$-parity have been kept open. 
Initial state radiation (ISR), final state radiation (FSR) and multiple 
interactions have been neglected. However, we take hadronization into account.
The final states studied here are :

\begin{itemize}
 \item Opposite sign dilepton ($OSD$) :
 $(\ell^{\pm}\ell^{\mp})+ (\geq 2)~ jets~ + \not{E_{T}}$ 
  
\item Same sign dilepton ($SSD$) : 
$(\ell^{\pm}\ell^{\pm})+ (\geq 2)~jets~ + ~~~~\not{E_{T}}$  

\item Single lepton $(1\ell+jets)$:  
$1\ell~ + (\geq 2)~ jets~ + \not{E_{T}}$ \footnote {Only for gaugino 
non-universality}   

\item Trilepton $(3\ell+jets)$: 
$3\ell~ + (\geq 2) ~jets~ + \not{E_{T}}$   

\item Hadronically quiet trilepton  $(3\ell)$:
$3\ell~ + \not{E_{T}}$   \footnote {Only for scalar 
non-universality}

\item Inclusive jets ($jets$): $(\geq 3) ~jets~ + X + \not{E_{T}}$    
\end{itemize}

\noindent
where $\ell$ stands for electrons and or muons. The cuts used 
are as follows \cite{ATLAS}:
(i) missing transverse energy $\not{E_{T}}$ $\geq ~100$ GeV, 
(ii) ${p_{T}}^l ~\ge ~20$ GeV and $|{{\eta}}_{\ell}| ~\le ~2.5$, (iii) 
an isolated lepton should have lepton-lepton separation
 ${\bigtriangleup R}_{\ell\ell}~ \geq 0.2$, lepton-jet separation 
 ${\bigtriangleup R}_{{\ell}j}~ \geq 0.4$, energy deposit ${E_{T}}$ 
due to jet activity around a lepton within 
$\bigtriangleup R~ \leq 0.2$ of the lepton axis should be $\leq 10$ GeV, 
(iv) ${E_{T}}^{jet} ~\geq ~100$ GeV and $|{\eta}_{jet}| ~\le ~2.5$ (absence of
such jets qualifies the event as hadronically quiet),  
(v) for the hadronically quiet trilepton events, we have used, in addition, 
invariant mass cut on the same flavour opposite
sign lepton pair as $|m_{Z} \sim M_{\ell_{+}\ell_{-}}| ~\geq 10$ GeV. 

We have generated all dominant standard model (SM) events  
in {\tt Pythia} for the same final states. The signal and 
background events have been all calculated for
an integrated luminosity of 300 fb$^{-1}$. 
As has been already mentioned, the ratios of events
in the different final states have been presented. Cases where any of the 
entries in the ratio has $\sigma=S/\sqrt{B}~\le~ 2$ 
($S$, $B$ being the numbers for signal and background events) 
have been specially marked with a '\#'.

\paragraph{{\bf{Results}:}} Here we provide some of the bar graphs in ratio 
space for gaugino and scalar non-universality discussed in our works 
\cite {Subho1,subho2} respectively. 
\begin{figure}[ht]
  \begin{tabular}{ll}
   \begin{minipage}[t]{3.5cm}
    \includegraphics[width=1.1\textwidth, height=.18\textheight]{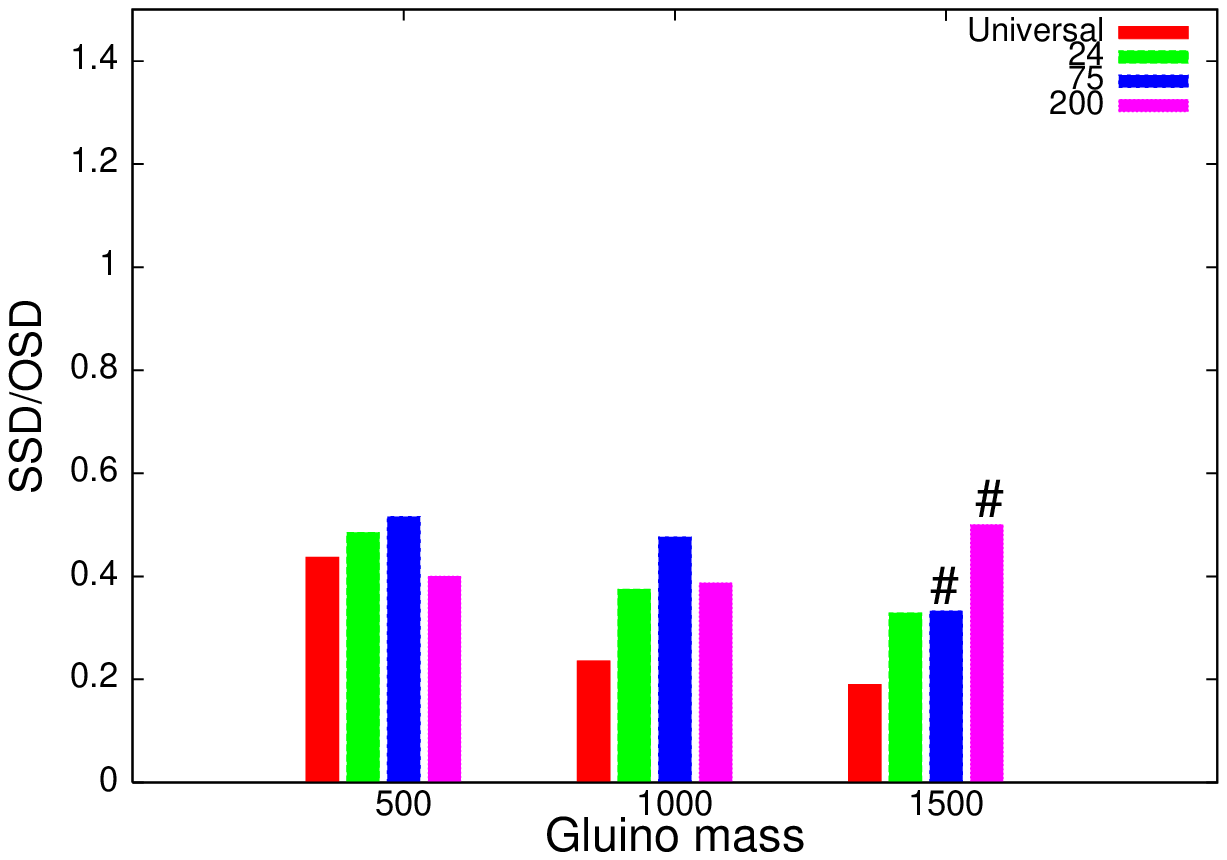}
   \end{minipage}
&
\begin{minipage}[t]{3.5cm}
    \includegraphics[width=1.1\textwidth,
    height=.18\textheight]{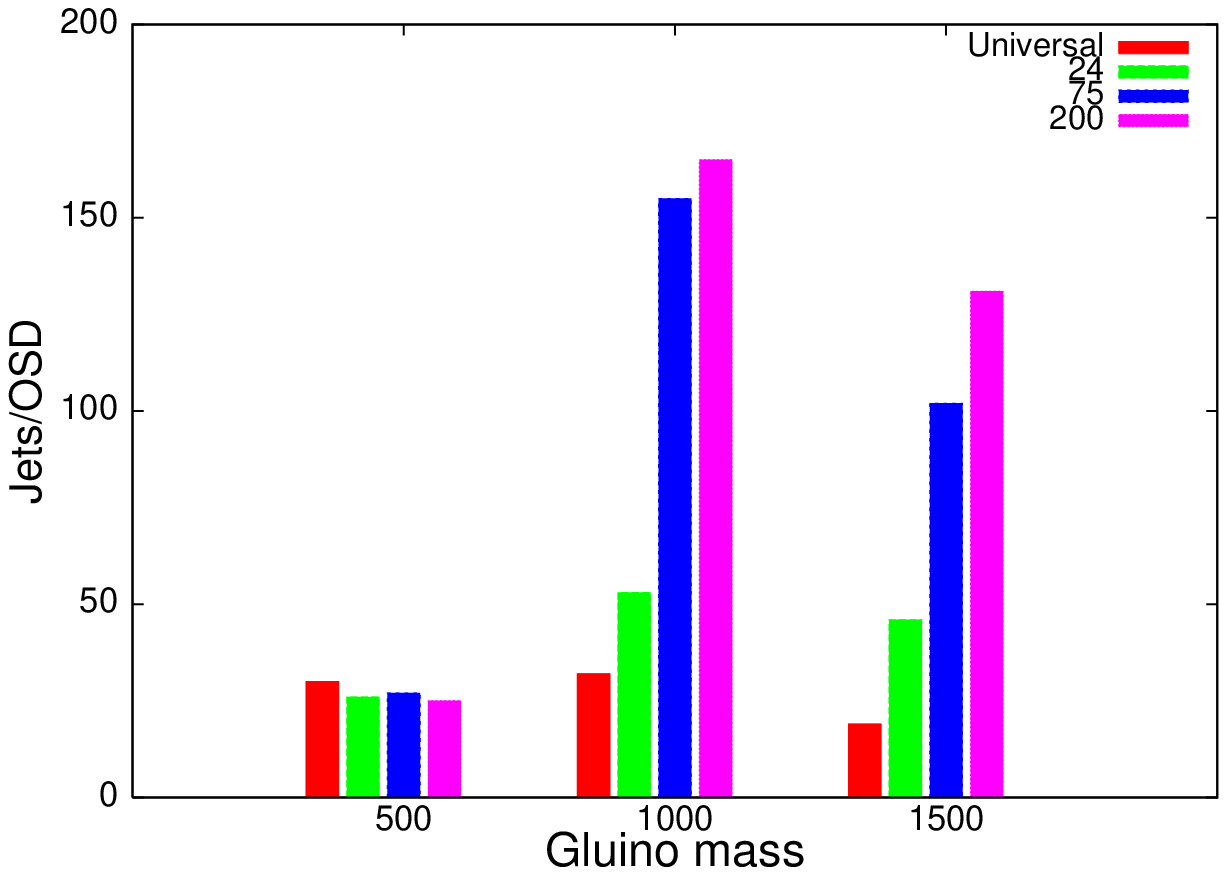}
   \end{minipage}
\end{tabular}
\caption{ \small SSD/OSD and JETS/OSD for gaugino-mass non-universality 
in SU(5): $m_{\tilde f}=$500 GeV, $\mu=$300 GeV, $\tan{\beta}=40$.}
\end{figure}
\begin{figure}[ht]
  \begin{tabular}{ll}
   \begin{minipage}[t]{3.5cm}
    \includegraphics[width=1.1\textwidth, height=.18\textheight]{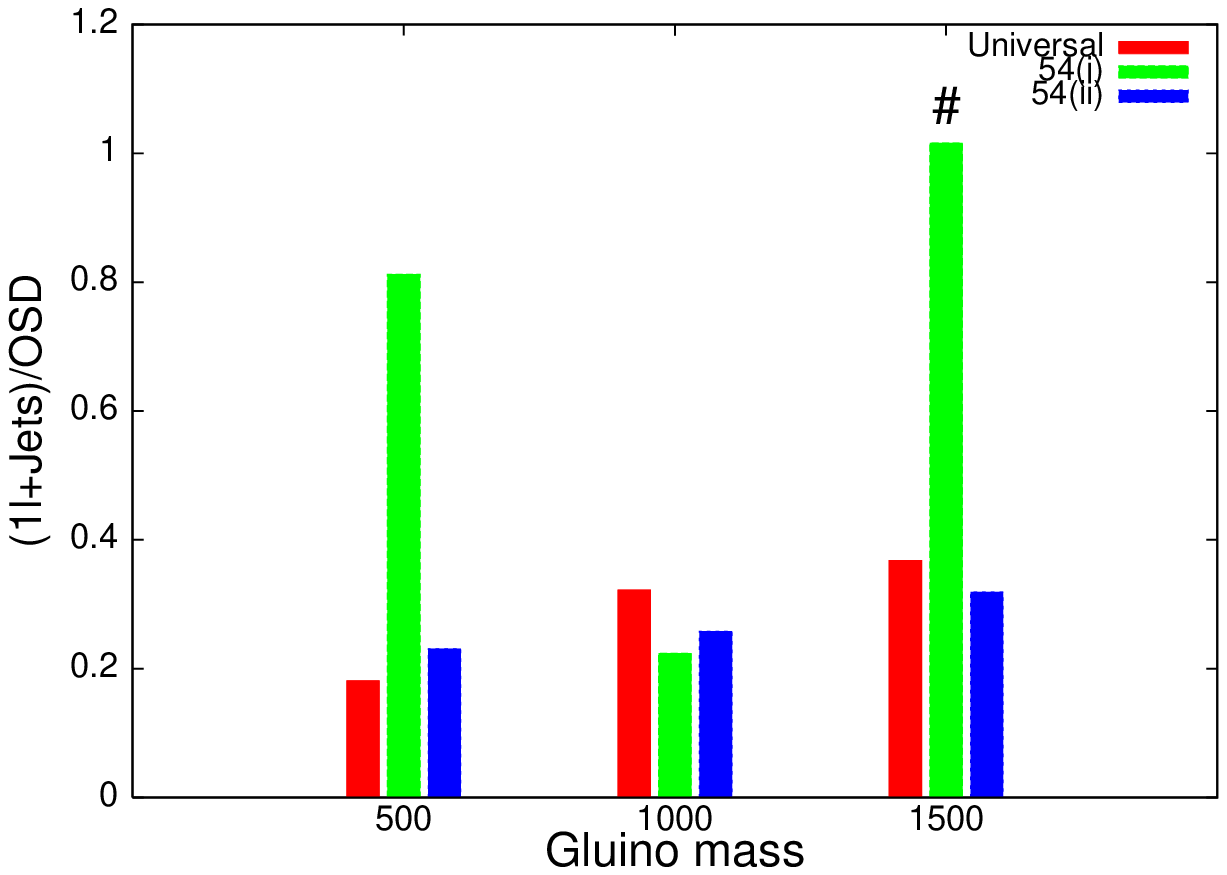}
   \end{minipage}
&
\begin{minipage}[t]{3.5cm}
    \includegraphics[width=1.1\textwidth,
    height=.18\textheight]{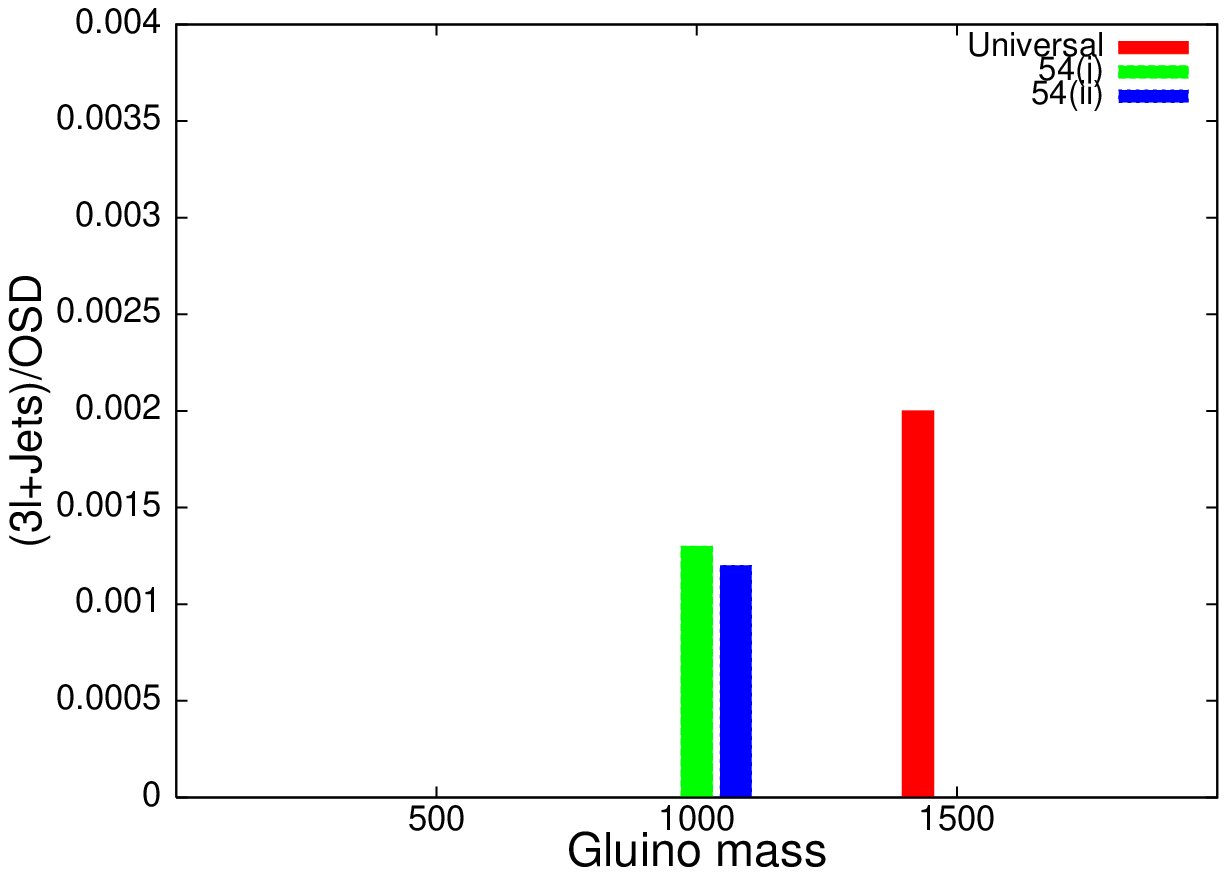} 
   \end{minipage}
\end{tabular}
\caption{ \small 1L+JETS/OSD and 3L+JETS/OSD for gaugino-mass non-universality 
in SO(10): $m_{\tilde f}=$1000 GeV, $\mu=$1000 GeV, $\tan{\beta}=5$.}
\end{figure}

Broadly, the features that emerge from the study of gaugino-mass 
non-universality (Figure 1 and 2) are as follows. In a substantial region 
of the parameter space ${\bf 75}$ and {\bf 200} of $SU(5)$ and 
${\bf 54 (i)}$  of $SO(10)$ are distinguishable. This 
is specifically because of the higher mass ratios of $M_1$ and $M_2$ with 
respect to $M_3$ in the above non-universal cases. However, 
distinguishing ${\bf 24}$ of $SU(5)$, ${\bf 54(ii)}$ of 
$SO(10)$ from each other and from the $universal~ case$ is 
relatively difficult. The trilepton channel turns out to be the most 
efficient discriminator in such cases.  

\begin{figure}[ht]
  \begin{tabular}{ll}
   \begin{minipage}[t]{3.5cm}
    \includegraphics[width=1.1\textwidth, height=.18\textheight]{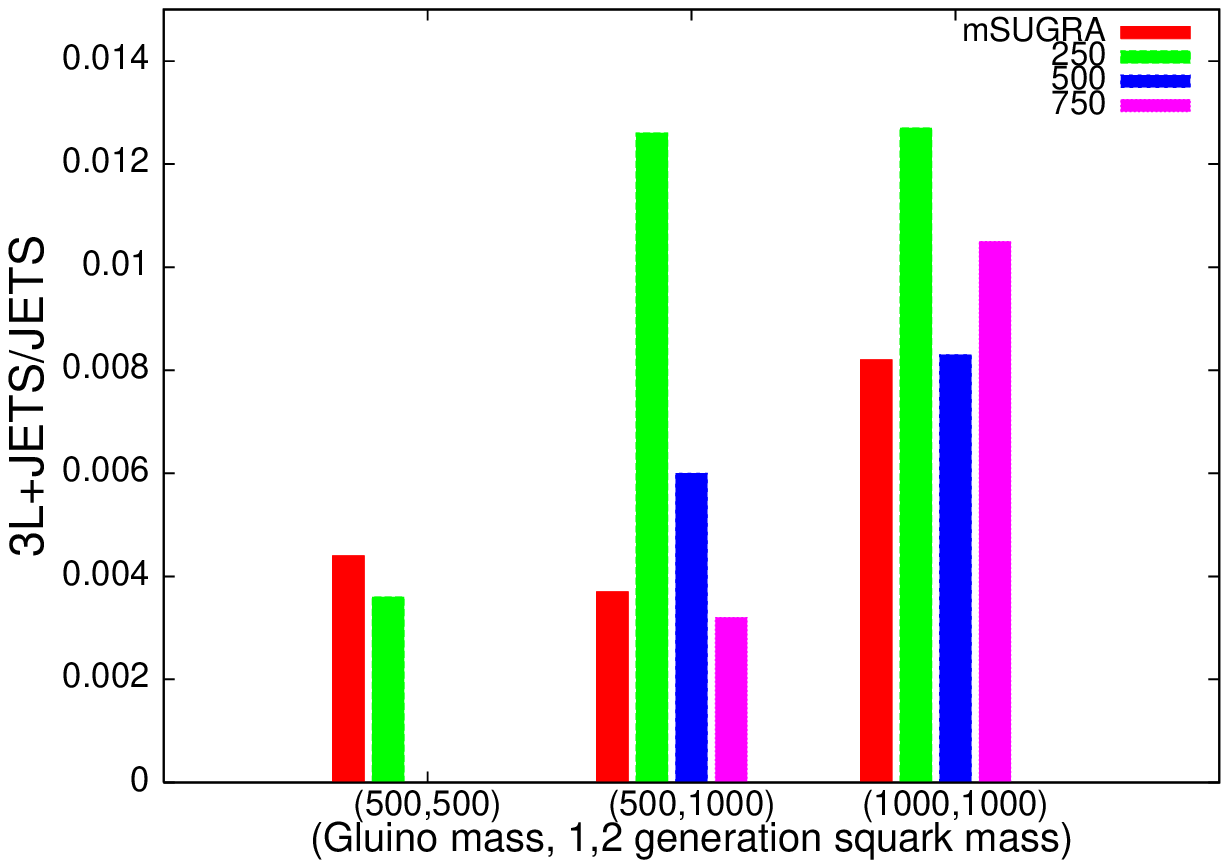}
   \end{minipage}
&
\begin{minipage}[t]{3.5cm}
    \includegraphics[width=1.1\textwidth,
    height=.18\textheight]{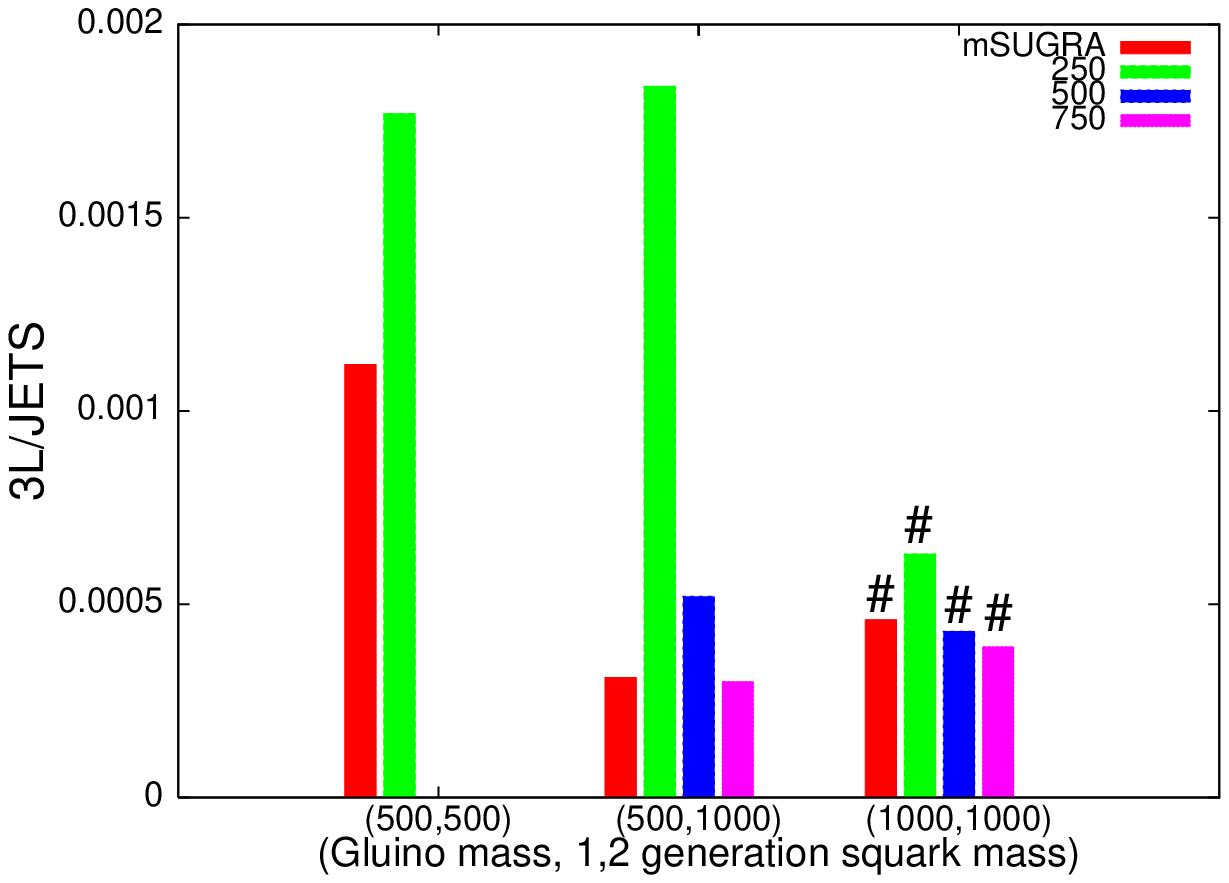} 
   \end{minipage}
\end{tabular}
\caption{ \small 3L+JETS/JETS and 3L/JETS ratios for Squark-slepton 
Non-universality, $\tan{\beta}=5$.}
\end{figure}

For squark-slepton non-universality (Figure 3), the case with 
$m_{\tilde l^{1,2}} = 250$ GeV (green ones), 
is fairly distinguishable from other cases as well as from 
mSUGRA especially for squark masses on the higher side.

\begin{figure}[ht]
  \begin{tabular}{ll}
   \begin{minipage}[t]{3.5cm}
    \includegraphics[width=1.1\textwidth, height=.18\textheight]{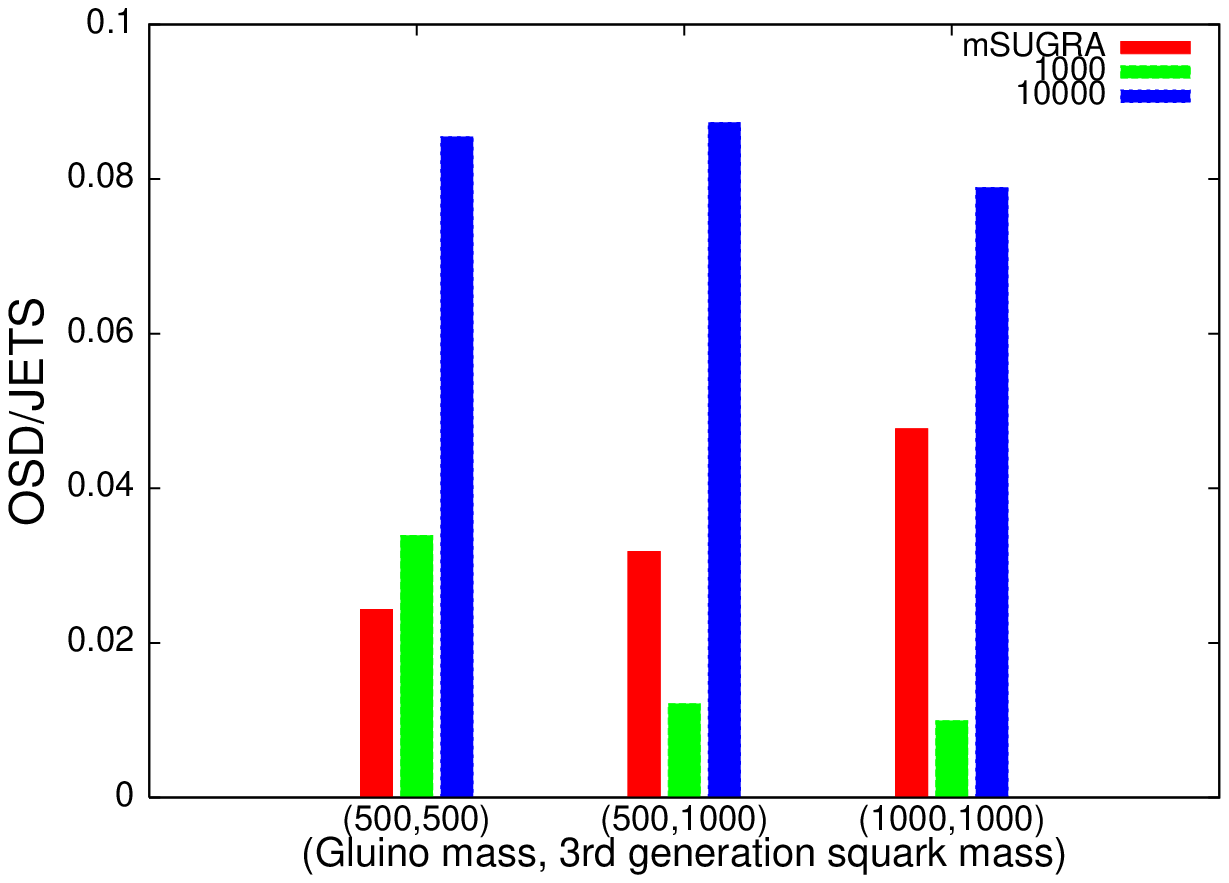}
   \end{minipage}
&
\begin{minipage}[t]{3.5cm}
    \includegraphics[width=1.1\textwidth,
    height=.18\textheight]{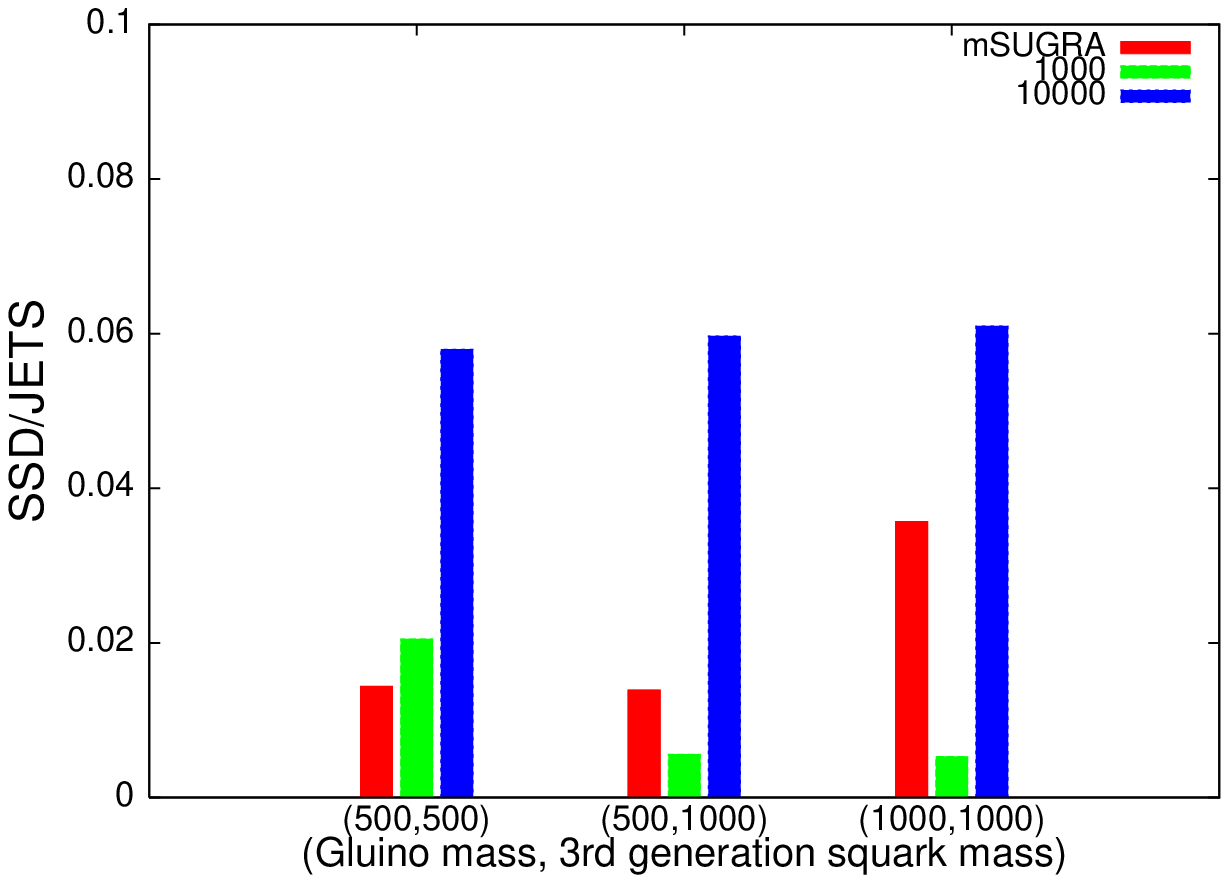} 
   \end{minipage}
\end{tabular}
\caption{ \small OSD/JETS and SSD/JETS ratios for third family scalar  
non-universality, $\tan{\beta}=5$.}
\end{figure}

In case of the third family scalar non-universality (Figure 4), when first 
two family squark masses are very high $m_{\tilde{q}^{1,2}} = 10$ TeV 
(blue ones), they can be easily identified from the rest of the scenarios 
as the leptonic branching fraction is higher in this case.

\begin{figure}[ht]
  \begin{tabular}{ll}
   \begin{minipage}[t]{3.5cm}
    \includegraphics[width=1.1\textwidth, height=.18\textheight]{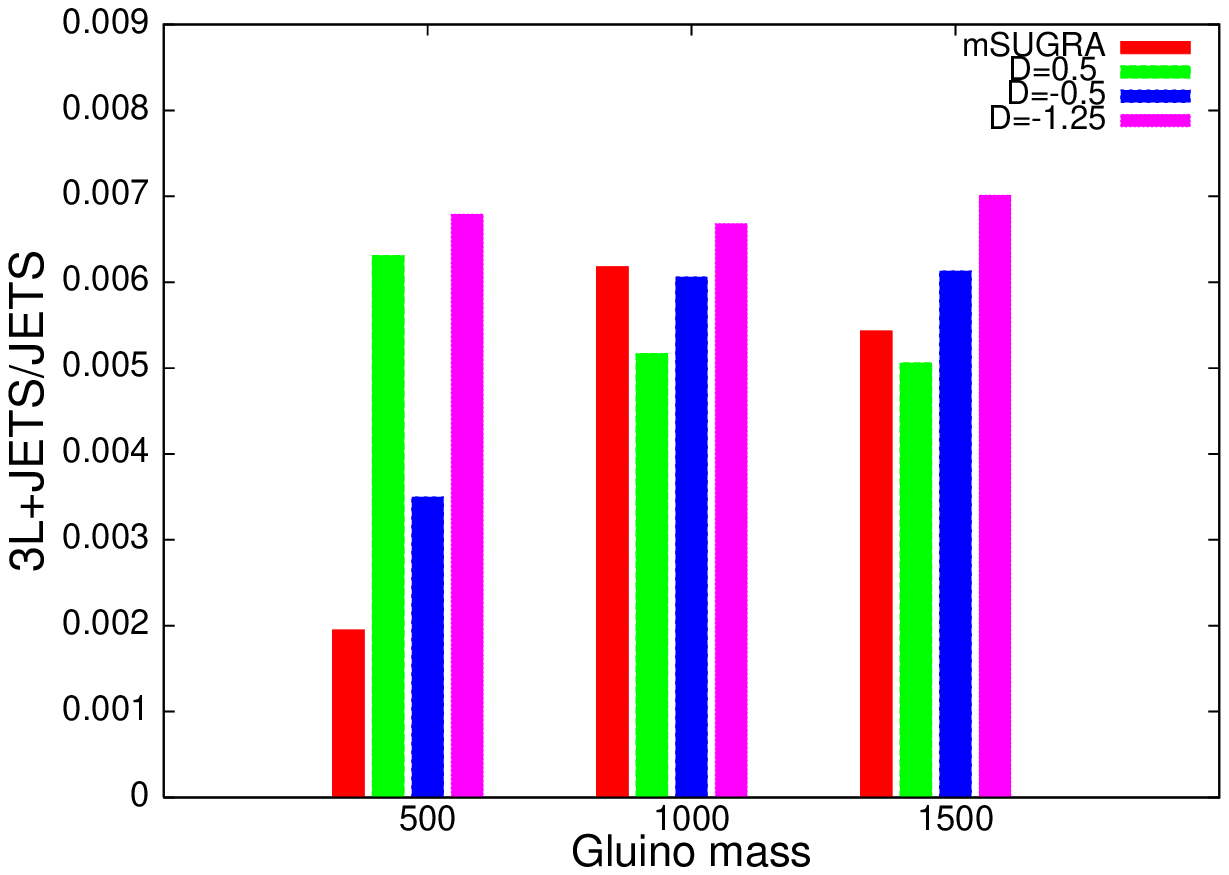}
   \end{minipage}
&
\begin{minipage}[t]{3.5cm}
    \includegraphics[width=1.1\textwidth,
    height=.18\textheight]{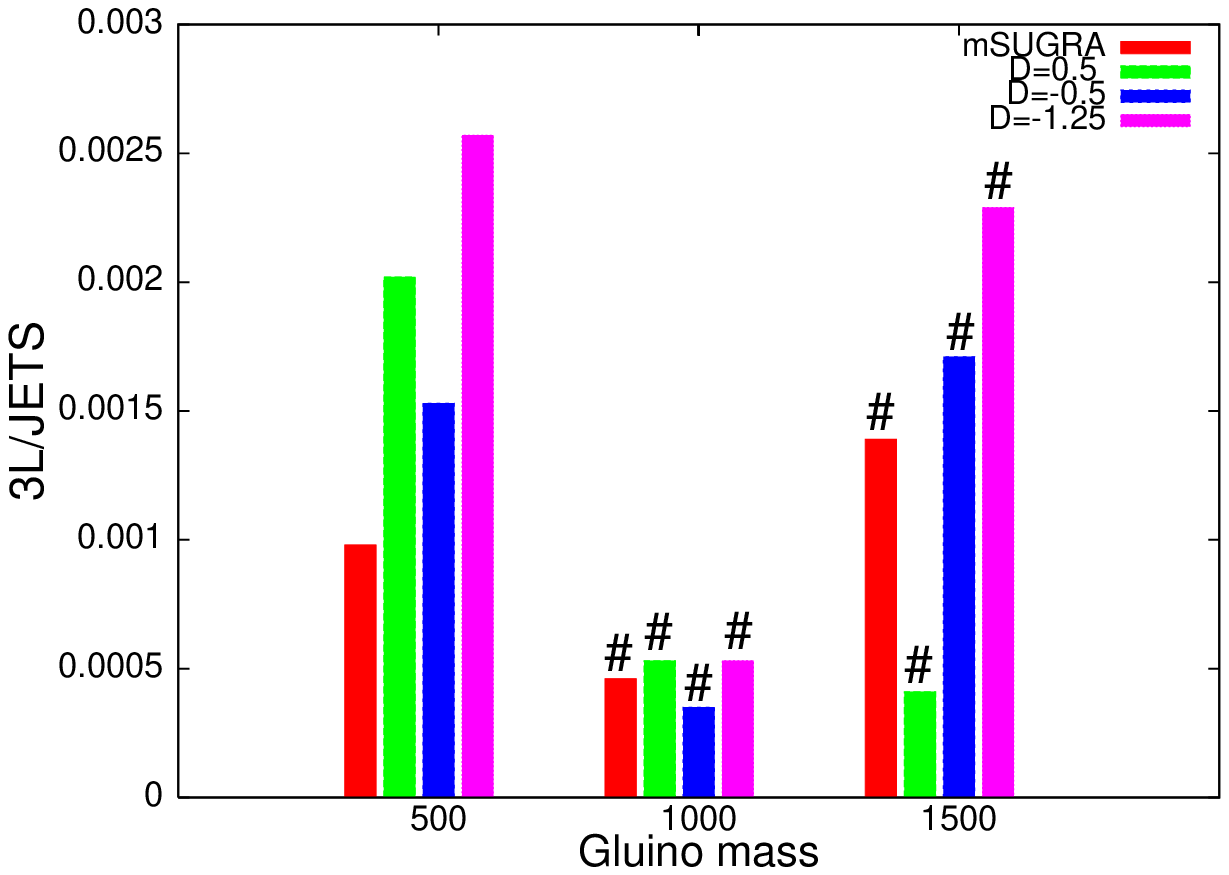} 
   \end{minipage}
\end{tabular}
\caption{ \small  3L+JETS/JETS and 3L/JETS ratios for non-universality 
coming from $SO(10)~D$-terms, $\tan{\beta}=5$.}
\end{figure}

In case of non-universality due to $SO(10)~D$-terms (Figure 5), for 
$m_{\tilde g}$= 500 GeV, $D$=0.5  and $D$=-1.25 
(green and purple ones respectively) are distinguishable 
from the ratios. However, it is difficult to distinguish various $D$-terms 
for higher gluino masses which is not so unexpected.

\paragraph{{\bf{Conclusions}:}} It is true that this investigation 
in terms of the scenarios chosen, is illustrative rather than being 
exhaustive. Still, the method advocated here can be useful in the so called 
`inverse problem' approach, where one aims to construct 
the underlying theory from a multichannel assortment of data.

\paragraph{{\bf Acknowledgements:}}
I would like to thank E. J. Chun for his support for attending SUSY08,
and AseshKrishna Datta and Biswarup Mukhopadhyaya for their collaboration. 
Our work was partially supported by funding available from the 
Department of Atomic Energy, 
Government of India, for the Regional Centre for Accelerator-based
Particle Physics, Harish-Chandra Research Institute.

\end{document}